\documentclass[journal=jacsat,manuscript=article, layout = twocolumn]{achemso}

\usepackage[version=3]{mhchem}
\usepackage[utf8]{inputenc}
\usepackage{mathtools}
\usepackage{graphicx}
\usepackage{subfigure}
\usepackage{float}
\usepackage{mathrsfs}

\usepackage{hyperref}
\usepackage{xcolor}
\usepackage{soul}
\usepackage[normalem]{ulem}

\newcommand\NAdeleted{\bgroup\markoverwith{\textcolor{blue}{\rule[0.5ex]{2pt}{0.4pt}}}\ULon}

\newcommand\RIdeleted{\bgroup\markoverwith{\textcolor{brown}{\rule[0.5ex]{2pt}{0.4pt}}}\ULon}

\graphicspath{ {images/} }

\newcommand{\angstrom}{\mbox{\normalfont\AA}}

\author{Reinis Irmejs}
\affiliation{St John's College, University of Cambridge}
\email{ri264@cam.ac.uk}
\author{Nadav Avidor}
\affiliation{Cavendish Laboratory, University of Cambridge, JJ Thomson Avenue, Cambridge CB30HE, United Kingdom}
\email{na364@cam.ac.uk}

\title[An \textsf{achemso} demo]{Atom Scattering For Chemical Analysis Of Surfaces}

\usepackage{natbib}
\usepackage{graphicx}

\begin{document}

\maketitle

\begin{abstract}
    The study explores machine learning methods for revealing chemical sensitivity in Helium spin-echo spectroscopy, in order to obtain ultra-sensitive surface analytic technique. We model bi-species co-adsorbed systems and demonstrate that by using deep-learning neural-networks partial surface concentrations are obtainable. An example system of particles with mass 50 and 100 a.m.u was tested with characteristic inter-adsorbate and adsorbate-substrate interactions, with partial surface concentrations being resolvable up to 20\% occupancy of adsorption sites, and with modestly high noise level of 4\%.
\end{abstract}

\section{Introduction}

Surface chemical reactions underpin processes applicable to water studies, catalysis, and atmospheric chemistry \cite{Klein798} \cite{Ma2015} \cite{wang2006} \cite{Wang2017} . Hence, they are the focus of extensive experimental and theoretical research. To monitor the progression of a reaction at a surface, the chemical composition (e.g. of co-adsorbed species) must be obtained. While several experimental methods have been pivotal in advancing the current understanding of surface chemistry \cite{BARTH200075} \cite{Gomer_1990} , certain experimental regimes remain largely inaccessible. For example, chemical analysis of a low-coverage adsorbate-surface (below 1\% adsorbates per surface sites) at temperatures of industrial relevance is considered nearly impossible. Furthermore, current scattering techniques for chemical analysis suffer from beam energies larger than the average chemical bond, that can result in excess perturbations to the surface, possibly biasing the analysis.

Measurements at low surface coverage are vital to experimentally separate adsorbate-substrate interactions from inter-adsorbate ones. In addition, measuring at elevated temperatures avoids the potential of freezing degrees of freedom, hence allow the studies of thermally activated reactions. Not only such results serve as an ultimate benchmark to \textit{ab-initio} calculations, they can provide entirely complementary information on reactions. 

Helium atom scattering can complement more traditional methods of chemical analysis of surfaces. It can measure at dilute surface coverages,temperatures as low as thousands of a kelvin, and with beam energies of a few milli-electronvolts. However, to-date helium scattering has not been used for chemical analysis due to difficulties in achieving chemical sensitivity. In this paper we combine computer simulations with machine learning methods to demonstrate that chemical analysis using helium scattering is achievable by applying Helium-3 Spin-Echo spectroscopy ($^3$HeSE), a technique for studying molecular dynamics at surfaces, at the sub-picosecond to nanosecond time-scale, with angstrom resolution in reciprocal space.

$^3$HeSE has been successfully applied to studies of surface diffusion of atoms and molecules \cite{JARDINE2009323}. $^3$HeSE experiment returns a signal that is directly proportional to the intermediate scattering function (ISF), which is itself a Fourier transform of the Van-Hove pair correlation function $G(\boldsymbol{R}, t)$. As such, the ISF contains a complete atomistic description of the surface \cite{VanHove1954249}. So far $^3$HeSE has not been used for multi-specie chemical analysis on surfaces, since the ability to resolve such line shapes was considered to be out of reach. 

We will now discuss the simulation of trajectories of adsorbate surface systems and calculation of the related ISFs and under the kinematic approximation. Following that, we will describe the use of Neural Networks (NN) for the analysing the ISF from a multi-species interacting adsorbate system, and the quantification of partial surface concentrations (number-densities) of the different species.

To simulate the adsorbate dynamics, we use PIGLE, a molecular dynamics simulator which solves the Langevin equation (1) for the particles \cite{avidor2019pigle}. The Langevin equation is written as

\begin{equation}
    m_{\mu} \ddot{\boldsymbol{r}}_{i} = - \nabla V_{\mu} -m \gamma_{\mu} \dot{\boldsymbol{r_{i}} }+ \boldsymbol{\xi}(t) + \sum_{i \neq j}{\boldsymbol{F}}_{i j}
\end{equation}

where the $\boldsymbol{r}_{i}$ denotes the trajectory of the $i^{th}$ particle, $\mu$ denoting the species of particles, $V$ the surface potential, $\gamma$ the drag coefficient, $ \boldsymbol{\xi}(t)$ the stochastic noise and $\boldsymbol{F}_{ij}$ the interactions between the particles \cite{avidor2019pigle}. 

Once the dynamics are solved for, the scattered amplitude from the i$^{th}$ adsorbate of the $\mu$'s species can be calculated assuming the kinematic approximation:
\begin{equation}
A_i(\Delta K,t) = F_{\mu}(\Delta K) \cdot e^{i\Delta K r_i(t)}
\end{equation}
with $F_{\mu}$ being the form-factor and the exponent being the phase factor of the amplitude scattered from the adsorbate. For a bi-species adsorbate system with species 1 and species 2, the total scattered amplitude at time 't' is given by $A(\Delta K,t) = \sum_i A_{1} + \sum_j A_{2}$, for 'i' and 'j' being the index for the scattering centers of species '1' and species '2' respectively. The coherent ISF can be calculated as the time auto-correlation of the scattered amplitudes, and is written as
\begin{equation}
\begin{aligned}
ISF(\Delta K,t) &= \mathscr{F}^{-1}\{\mathscr{F}\{A^*\} \cdot \mathscr{F}\{A\}\} \\ &= \mathscr{F}^{-1}\{S(\Delta K,\Delta \omega)\}
\end{aligned}
\end{equation}

$\mathrm{S}$ is the scattering function, the quantity which is being measured at a helium time-of-flight experiment. For a bi-species system, $\mathrm{S}$ is given by
\begin{equation}
\begin{aligned}
&S(\Delta K,t) = \\
&(\mathscr{F}\{A_1^*\} + \mathscr{F}\{A_2^*\}\} \cdot \{ \mathscr{F}\{A_1\} + \mathscr{F}\{A_2\} )\\
&= S_1 + S_2 \\ &+ \mathscr{F}\{A_2^*\} \cdot \mathscr{F}\{A_1\} + \mathscr{F}\{A_1^*\} \cdot \mathscr{F}\{A_2\}
\end{aligned}
\end{equation}

Therefore, the ISF of a bi-species system is equal to the linear combination of the single-species ISFs, plus mixed-terms which are governed by the correlation between the two species. When the correlation between these two species is negligible, such as in cases of weak inter-species interactions or a sparse adsorbate systems, the mixed-terms vanish. We now seek to determine, whether a bi-species system can be quantified, with prior knowledge only on ISF$_1$ and ISF$_2$. In other words, can we extract the relative concentration of species 1 \& 2 in a co-adsorbed system (with unknown relative concentrations), if we have previously calculated both species separately.

\subsection{Deep-learning network structure}

Extracting information about the relative concentrations from the separately calculated ISF data is not possible analytically, hence we created a deep learning model to classify the relative concentrations. Since ISFs for different $\Delta K$ values form 2D data, convolutional neural networks (CNNs) appeared as a good deep-learning technique to use for classification as they have been widely used in image recognition and time-series classification problems \cite{fawaz2019deep} due to their ability to extract and classify features in the data. Similarly to images, the ISFs have a range of features describing the adsorbate dynamics, like oscillations in the early ISF stages (short term behaviour) and the decay rate of the ISF (long term behaviour). Use of a recurennt neural network (LSTM) was also explored as it is widely used in sequence classification, although its accuracy proved to be lower than CNN's. The CNN that was used had a 2D input of 5 $\Delta K$ values $0.5$, $1.0$, $2.5$, $3.0$, $4.0$ $\angstrom^{-1}$ and 1025 time steps of ISF for a total decay time of $25$ $ps$. The network consists of 2 sets of convolution and pooling layers followed by a densely connected layer and a classification layer that gives the relative concentrations of each species. 

\subsection{Data pre-processing}

To increase the prediction accuracy, ISFs were normalised from 0 to 1, and the relative concentrations were used as training labels. To better resemble the data obtained experimentally, a white Gaussian noise of a certain amplitude was added to the data. In the results section noise amplitude is given as a fraction of the maximum ISF value in percent. 

Throughout the analysis, the total and each adsorbate concentration values were measured in terms of the fraction of the adsorption sites filled. 
The training labels were also normalised to give relative concentrations:

\begin{equation}
\centering
    RC_{\mu} = \frac{C_{\mu}}{C_{A}+C_{B}}
\end{equation}

where $C_{\mu}$ and $RC_{\mu}$ are the absolute concentration and the relative concentration values of species $\mu$.

However, to better interpret the results, the values were converted back to the absolute values in the next section. 

The prediction accuracy was chosen to be the root mean squared error (RMSE) (6).

\begin{equation}
\centering
    RMSE = \frac{{\sqrt{\sum{(C_{pred} - C_{test})^2}}}}{N}
\end{equation}

\section{Results and Discussion}

It is important to distinguish between the two sets of data that were used: linearly added and complete. The linearly added data was created by linearly adding separately simulated ISFs of species $A$ and $B$ scaled by their concentrations. The complete data was created by simulating both species together. To check whether CNN can perform such a classification, the model was first trained and tested on the linearly added data. The CNN model proved to be successful in classifying the partial concentrations with a relative $RMSE ~\sim 2.5 \%$ (included in figure 4). This accuracy is high enough to distinguish the ratios of species in chemical reactions, with the relative error in the ratio of 5 \%. 
However, such linearly added ISFs does not exist in practice, as inter-specie interactions are neglected when separate ISFs are linearly combined. 
Using deep learning for classification requires a considerable dataset of order $10^3$ examples at various different concentrations of species $A$ and $B$. To experimentally obtain $\sim1000$ complete ISF signals at various different concentrations, it would be necessary to perform the same number of $^3$HeSE experiments. This is very time consuming in practice. However, if linearly added data of species $A$ and $B$ is used, it suffices to only perform $\sim50$ experiments for each specie to make a dataset of $\sim 2.5 * 10^3$ examples with only performing $\sim100$ experiments in total. Hence, to make the method feasible experimentally, sufficient accuracy must be obtained using the linearly added data. 
Furthermore, if one would be interested in distinguishing ratios of $A$ and another specie $C$, it would only be necessary to run new experiments for $C$ and linearly combine with already existing results from $A$. Future work may involve classifying signals of each molecule of interest with high accuracy and creating a library that could be used to distinguish between any molecular species of interest.

Further, the prediction results for data of species $A$ and $B$ with both intra and inter-specie interactions are analysed with a network trained on the linearly added ISFs. Benchmark precision values are obtained by testing the model on linearly added data. 

\subsection{Analysis of data with both intra and inter-specie interactions }

In this subsection, prediction accuracy in terms of RMSE is given for various added Gaussian noise values $(0\%, 2\%, 4\%, 6\%, 8\%)$. Currently, the experimental noise is $\sim 1\%$ \cite{Townsend2018thesis}, thus the slightly larger $2$ $\%$ was chosen to be added to the training data. This value is important as, the CNN model proved to perform the best on test data with the same added noise level. Figure 1 gives an example of how the accuracy changes with total concentration for test data with $2\%$ noise and also includes the error values for each data point to better assess the distribution of the errors. 

\begin{figure}
    \centering
    \includegraphics[width = \linewidth]{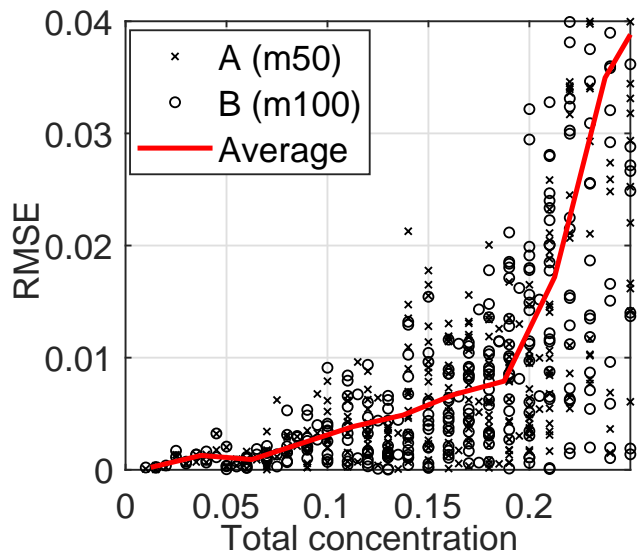}
    \caption{Figure shows how the prediction accuracy varies with the total concentration. The data points of each prediction are included to better assess the error spread. It can be seen that almost no points have higher than $2 \sigma$ error. The prediction accuracy increases linearly up until concentration 0.2 after which it rises abruptly.}
\end{figure}

Results from figure 1 yield that the averaged RMSE error very well represents the overall picture, hence further only the averaged error will be included. In figure 2 we assess how the prediction accuracy varies with added noise for the data with inter-specie interaction of strength $75 * 10^3 a.u.$ (similar to $H$ and $CO$ interaction strength).

\begin{figure}
    \centering
    \includegraphics[width = \linewidth]{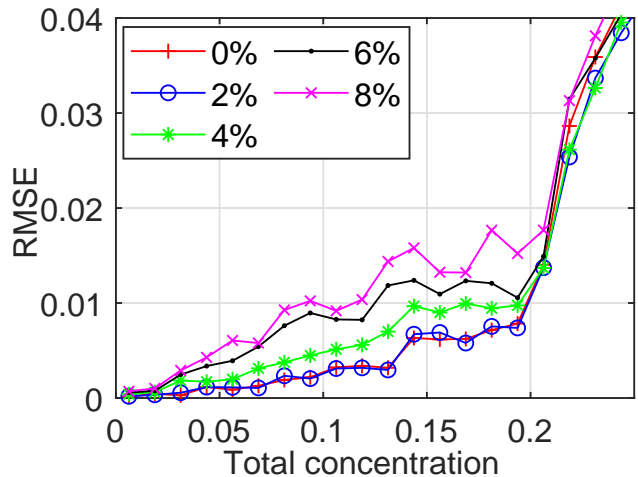}
    \caption{Figure shows how the prediction accuracy varies with the total concentration for various noise levels. For the low noise values of 0\% and 2\% the errors are almost identical and rises linearly with relative prediction accuracy of ~3.5\%. The inter-specie interaction value here is $75*10^3 a.u.$)}
\end{figure}

From figure 3, it can be seen that no matter what the noise level, the prediction accuracy jump at around concentration $0.2$ indicating that that is where the low inter-specie interaction regime completely breaks down. However, for a larger inter-specie interaction the jump in $RMSE$ might happen at different occupation value. To explore this, the same analysis was done on data with twice the inter-specie interaction value (figure 3). The $RMSE$ values for both interaction strengths are compared with benchmark accuracy in figure 4.

\begin{figure}
    \centering
    \includegraphics[width = \linewidth]{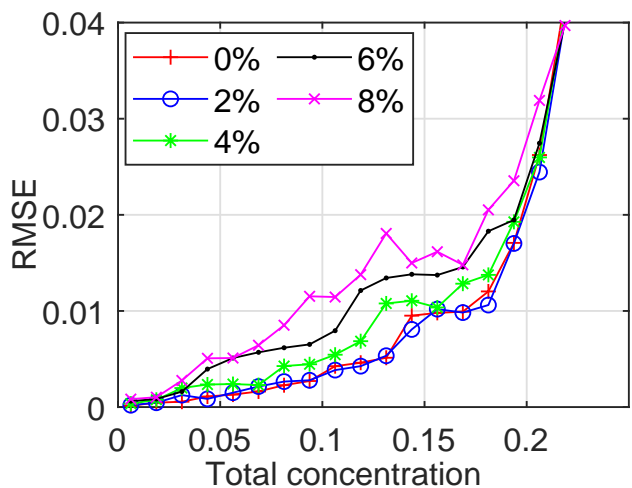}
    \caption{Figure shows how the prediction accuracy varies with the total concentration for various noise levels. For the low noise values of 0\% and 2\% the errors are almost identical and rises linearly up until concentration $0.13$ with relative prediction accuracy of ~3.5\%. The inter-specie interaction value here is $150*10^3 a.u.$}
\end{figure}

\begin{figure*}
    \centering
    \includegraphics[width = 0.7\linewidth]{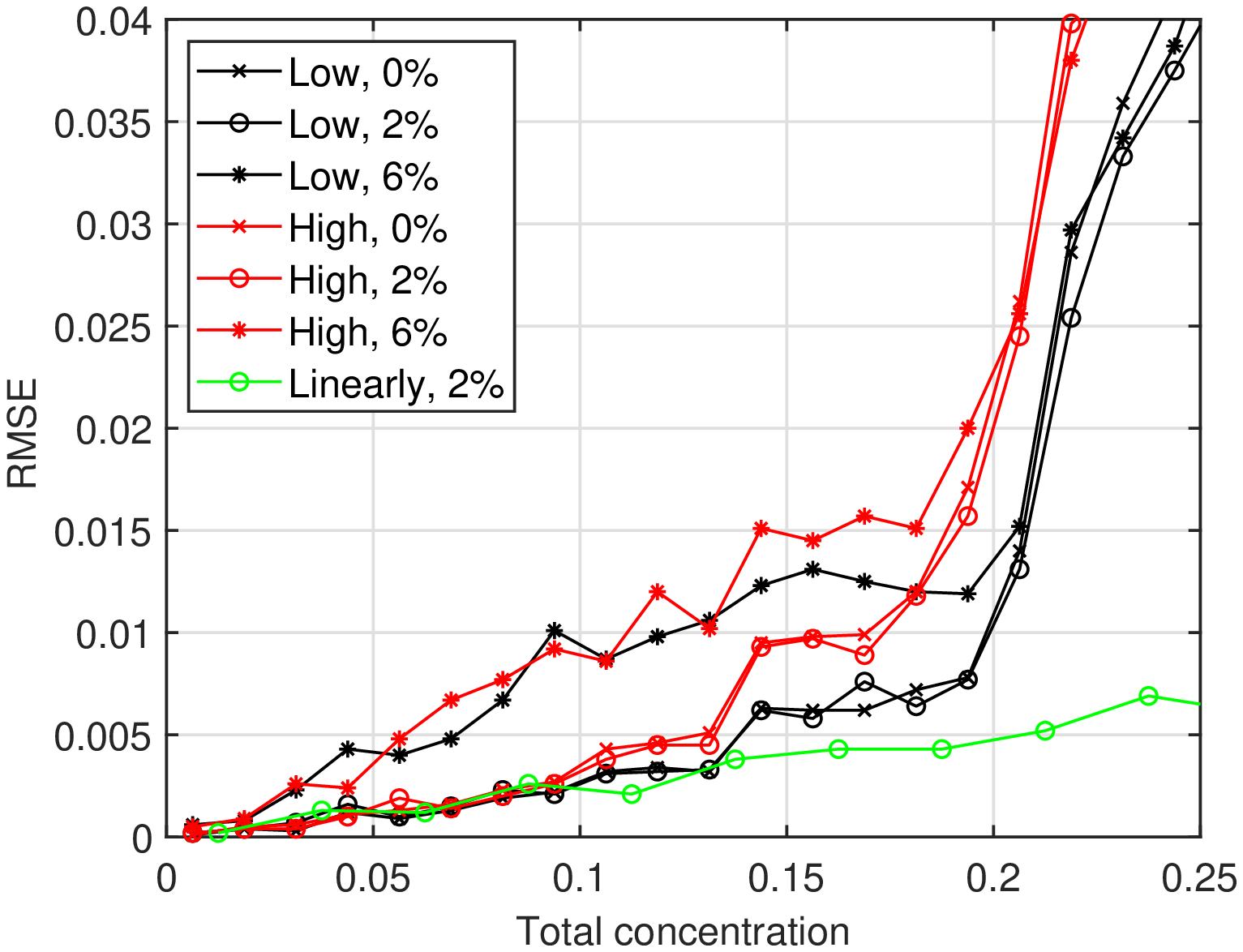}
    \caption{Figure compares the prediction accuracy for both inter-specie interaction strengths with the benchmark values. Up until concentration of 0.1 there is no noticeable difference in the accuracy for all 3 cases, after which higher - interaction data does worse. }
\end{figure*}

Figure 4 shows no noticeable impact in prediction accuracy from inter-specie interactions below concentration of $0.1$. After that, the lower interaction diverges away from the no - interaction value slower than the higher, confirming that the inter-specie interaction is what causes the model to break down. After concentration of $0.2$, the model breaks down and the $RMSE$ error jumps for data with inter-specie interactions. 
\section{Conclusion}

The results show that the deep-learning model is able to successfully predict the relative concentrations for each specie with accuracy of $2.5\%$ in the low inter-specie interaction regime for fractional adsorption site occupation up to $10 \%$ with no difference in accuracy from the benchmark values (figure 4). In the mid interaction range for fractional occupation between $0.1$ and $0.2$, the model is still able to reliably predict the relative concentrations, although the error is dependent on the interaction strength (figure 5). Figures 3 \& 4  show that the prediction accuracy is $<5\%$ for an added Gaussian noise levels of up to $4\%$. 

The fractional occupation and noise levels are well within the range of ones used in $^3$HeSE, with the latter being of $\sim 1 \%$ \cite{Townsend2018thesis}. Furthermore, with more data it might be possible to further increase the accuracy and acceptable noise levels.
Therefore, in theory it should be possible to adapt the method for experimentally obtained data and analytically investigate surface chemical reactions. Further work must be done to compile enough experimental data to train such a CNN model. If successful, this method could yield brand-new insights in the surface chemistry. 

\section{Acknowledgements}
The work was conducted as part of the Undergraduate Summer Research Program, ‘Surface nanoPhysics and Atom-Surface Scattering’, provided by the Cambridge Atom Scatting Facility(CASF) at the Cavendish Laboratory, University of Cambridge. The Engineering and Physical Sciences Research Council (EPSRC) is acknowledged for the financial support to CASF(EP/T00634X/1).  N. A. gratefully acknowledges the Herchel Smith for funding. This work has been performed using resources provided by the Cambridge Tier-2 system operated by the University of Cambridge Research Computing Service (www.hpc.cam.ac.uk) funded by EPSRC Tier-2 capital grant EP/P020259/1.

\bibliography{references}

\providecommand{\latin}[1]{#1}
\makeatletter
\providecommand{\doi}
  {\begingroup\let\do\@makeother\dospecials
  \catcode`\{=1 \catcode`\}=2 \doi@aux}
\providecommand{\doi@aux}[1]{\endgroup\texttt{#1}}
\makeatother
\providecommand*\mcitethebibliography{\thebibliography}
\csname @ifundefined\endcsname{endmcitethebibliography}
  {\let\endmcitethebibliography\endthebibliography}{}
\begin{mcitethebibliography}{12}
\providecommand*\natexlab[1]{#1}
\providecommand*\mciteSetBstSublistMode[1]{}
\providecommand*\mciteSetBstMaxWidthForm[2]{}
\providecommand*\mciteBstWouldAddEndPuncttrue
  {\def\EndOfBibitem{\unskip.}}
\providecommand*\mciteBstWouldAddEndPunctfalse
  {\let\EndOfBibitem\relax}
\providecommand*\mciteSetBstMidEndSepPunct[3]{}
\providecommand*\mciteSetBstSublistLabelBeginEnd[3]{}
\providecommand*\EndOfBibitem{}
\mciteSetBstSublistMode{f}
\mciteSetBstMaxWidthForm{subitem}{(\alph{mcitesubitemcount})}
\mciteSetBstSublistLabelBeginEnd
  {\mcitemaxwidthsubitemform\space}
  {\relax}
  {\relax}

\bibitem[Klein and Shinoda(2008)Klein, and Shinoda]{Klein798}
Klein,~M.~L.; Shinoda,~W. Large-Scale Molecular Dynamics Simulations of
  Self-Assembling Systems. \emph{Science} \textbf{2008}, \emph{321},
  798--800\relax
\mciteBstWouldAddEndPuncttrue
\mciteSetBstMidEndSepPunct{\mcitedefaultmidpunct}
{\mcitedefaultendpunct}{\mcitedefaultseppunct}\relax
\EndOfBibitem
\bibitem[Ma \latin{et~al.}(2015)Ma, Grey, Shen, Urbakh, Wu, Liu, Liu, and
  Zheng]{Ma2015}
Ma,~M.; Grey,~F.; Shen,~L.; Urbakh,~M.; Wu,~S.; Liu,~J.~Z.; Liu,~Y.; Zheng,~Q.
  Water transport inside carbon nanotubes mediated by phonon-induced
  oscillating friction. \emph{Nature Nanotechnology} \textbf{2015}, \emph{10},
  692--695\relax
\mciteBstWouldAddEndPuncttrue
\mciteSetBstMidEndSepPunct{\mcitedefaultmidpunct}
{\mcitedefaultendpunct}{\mcitedefaultseppunct}\relax
\EndOfBibitem
\bibitem[Wang \latin{et~al.}(2006)Wang, Li, and Xia]{wang2006}
Wang,~Y.-G.; Li,~H.-Q.; Xia,~Y.-Y. Ordered Whiskerlike Polyaniline Grown on the
  Surface of Mesoporous Carbon and Its Electrochemical Capacitance Performance.
  \emph{Advanced Materials} \textbf{2006}, \emph{18}, 2619--2623\relax
\mciteBstWouldAddEndPuncttrue
\mciteSetBstMidEndSepPunct{\mcitedefaultmidpunct}
{\mcitedefaultendpunct}{\mcitedefaultseppunct}\relax
\EndOfBibitem
\bibitem[Wang \latin{et~al.}(2017)Wang, R{\"u}hling, Amirjalayer, Knor, Ernst,
  Richter, Gao, Timmer, Gao, Doltsinis, Glorius, and Fuchs]{Wang2017}
Wang,~G.; R{\"u}hling,~A.; Amirjalayer,~S.; Knor,~M.; Ernst,~J.~B.;
  Richter,~C.; Gao,~H.-J.; Timmer,~A.; Gao,~H.-Y.; Doltsinis,~N.~L.;
  Glorius,~F.; Fuchs,~H. Ballbot-type motion of N-heterocyclic carbenes on gold
  surfaces. \emph{Nature Chemistry} \textbf{2017}, \emph{9}, 152--156\relax
\mciteBstWouldAddEndPuncttrue
\mciteSetBstMidEndSepPunct{\mcitedefaultmidpunct}
{\mcitedefaultendpunct}{\mcitedefaultseppunct}\relax
\EndOfBibitem
\bibitem[Barth(2000)]{BARTH200075}
Barth,~J. Transport of adsorbates at metal surfaces: from thermal migration to
  hot precursors. \emph{Surface Science Reports} \textbf{2000}, \emph{40}, 75
  -- 149\relax
\mciteBstWouldAddEndPuncttrue
\mciteSetBstMidEndSepPunct{\mcitedefaultmidpunct}
{\mcitedefaultendpunct}{\mcitedefaultseppunct}\relax
\EndOfBibitem
\bibitem[Gomer(1990)]{Gomer_1990}
Gomer,~R. Diffusion of adsorbates on metal surfaces. \emph{Reports on Progress
  in Physics} \textbf{1990}, \emph{53}, 917--1002\relax
\mciteBstWouldAddEndPuncttrue
\mciteSetBstMidEndSepPunct{\mcitedefaultmidpunct}
{\mcitedefaultendpunct}{\mcitedefaultseppunct}\relax
\EndOfBibitem
\bibitem[Jardine \latin{et~al.}(2009)Jardine, Hedgeland, Alexandrowicz,
  Allison, and Ellis]{JARDINE2009323}
Jardine,~A.; Hedgeland,~H.; Alexandrowicz,~G.; Allison,~W.; Ellis,~J. Helium-3
  spin-echo: Principles and application to dynamics at surfaces. \emph{Progress
  in Surface Science} \textbf{2009}, \emph{84}, 323 -- 379\relax
\mciteBstWouldAddEndPuncttrue
\mciteSetBstMidEndSepPunct{\mcitedefaultmidpunct}
{\mcitedefaultendpunct}{\mcitedefaultseppunct}\relax
\EndOfBibitem
\bibitem[Van~Hove(1954)]{VanHove1954249}
Van~Hove,~L. Correlations in space and time and born approximation scattering
  in systems of interacting particles. \emph{Physical Review} \textbf{1954},
  \emph{95}, 249--262, cited By 1695\relax
\mciteBstWouldAddEndPuncttrue
\mciteSetBstMidEndSepPunct{\mcitedefaultmidpunct}
{\mcitedefaultendpunct}{\mcitedefaultseppunct}\relax
\EndOfBibitem
\bibitem[Avidor \latin{et~al.}(2019)Avidor, Townsend, Ward, Jardine, Ellis, and
  Allison]{avidor2019pigle}
Avidor,~N.; Townsend,~P.~S.; Ward,~D.; Jardine,~A.; Ellis,~J.; Allison,~W.
  PIGLE—Particles Interacting in Generalized Langevin Equation simulator.
  \emph{Computer Physics Communications} \textbf{2019}, \emph{242},
  145--152\relax
\mciteBstWouldAddEndPuncttrue
\mciteSetBstMidEndSepPunct{\mcitedefaultmidpunct}
{\mcitedefaultendpunct}{\mcitedefaultseppunct}\relax
\EndOfBibitem
\bibitem[Fawaz \latin{et~al.}(2019)Fawaz, Forestier, Weber, Idoumghar, and
  Muller]{fawaz2019deep}
Fawaz,~H.~I.; Forestier,~G.; Weber,~J.; Idoumghar,~L.; Muller,~P.-A. Deep
  learning for time series classification: a review. \emph{Data Mining and
  Knowledge Discovery} \textbf{2019}, \emph{33}, 917--963\relax
\mciteBstWouldAddEndPuncttrue
\mciteSetBstMidEndSepPunct{\mcitedefaultmidpunct}
{\mcitedefaultendpunct}{\mcitedefaultseppunct}\relax
\EndOfBibitem
\bibitem[Townsend(2018)]{Townsend2018thesis}
Townsend,~P.~S. Diffusion of light adsorbates on transition metal surfaces.
  Ph.D.\ thesis, University of Cambridge, 2018\relax
\mciteBstWouldAddEndPuncttrue
\mciteSetBstMidEndSepPunct{\mcitedefaultmidpunct}
{\mcitedefaultendpunct}{\mcitedefaultseppunct}\relax
\EndOfBibitem
\end{mcitethebibliography}
\end{document}